\documentclass[%
mathleft,%
]{an}
\usepackage{graphicx}
\usepackage{epstopdf}
\usepackage{times}
\usepackage{float}
\usepackage{textcomp}
\usepackage{natbib}
\usepackage{hyperref}

\begin{document}

\Pagespan{1}{}
\Yearpublication{2015}%
\Yearsubmission{2015}%
\Month{1}%
\Volume{999}%
\Issue{92}%

\title{Hamburger Sternwarte plate archives: Historic long-term variability study
of active galaxies based on digitized photographic plates}

\author{ M.\,Wertz\inst{1,2}, D.\,Horns\inst{1}, D.\,Groote\inst{2}, T.\,Tuvikene\inst{3,4}, S.\,Czesla\inst{2}, 
J.H.M.M.\,Schmitt\inst{2} 
}
\titlerunning{}

\authorrunning { M.\,Wertz, D.\,Horns, D.\,Groote, T.\,Tuvikene, S.\,Czesla, J.H.M.M.\,Schmitt}

\institute{Institut f\"ur Experimentalphysik,
Universit\"at Hamburg,
Germany
\and
Hamburger Sternwarte,
Universit\"at Hamburg,
Germany
\and
Leibnitz-Institut f\"ur Astrophysik Potsdam, Germany
\and
Tartu Observatory, Estonia
}

\received{XXXX}
\accepted{XXXX}
\publonline{XXXX}


\abstract{
At the Hamburger Sternwarte an effort was started in 2010 with the aim of digitizing its more 
than 45\,000 photographic plates and films stored in its plate archives. At the time of writing, 
more than 31\,000 plates have already been made available on the Internet for researchers, 
historians, as well as for the interested public. The digitization process and the Internet 
presentation of the plates and accompanying hand written material (plate envelopes, logbooks, 
observer notes) are presented here. To fully exploit the unique photometric and astrometric data, 
stored on the plates, further processing steps are required including registering the plate to 
celestial coordinates, masking of the plates, and a calibration of the photo-emulsion 
darkening curve. To demonstrate the correct functioning of these procedures, historical light 
curves of two bright BL Lac type active galactic nuclei are extracted. The resulting light curve 
of the blazar 1ES~1215+303 exhibits a large decrease in the magnitude from 
$14.25^{+0.07}_{-0.12}$ to $15.94^{+0.09}_{-0.13}$ in about 300~days, which proves the 
variability in the optical region. Furthermore, we compare the measured magnitudes for the 
quasar 3C~273 with contemporaneous measurements.
}

\maketitle

\section{Introduction}

The advent of photographic emulsion techniques to detect and store telescopic observations in the 
second half of the 19$^{th}$ century provided a challenge for contemporaneous data 
analysis techniques. Most of the astronomical objects captured by the plate emulsion
were not available for scientific analysis and were essentially ignored, given the rather 
tedious techniques required to extract astrometric and photometric information on individual 
objects.

There are two major problems with photographic plates: First, the information 
contained on the plates is not readily available in digital form, and second, the 
photographic emulsions have a finite lifetime and thus all information contained on any 
given plate will eventually be lost, since there are no backup plates. The latter 
issue is particularly relevant since the lifetime of emulsions is around
one hundred years.

In the current era of CCD imaging, powerful techniques have 
been developed by astronomers to extract digital data from CCD frames using the enormous 
processing power of present-day digital computers. Thus, it is a natural step to try to 
recover the unique information captured by historical observations using modern data 
analysis methods. 

In order to accomplish that goal, the information contained in the plate emulsion needs 
to be converted into some digital form with the help of a scanning process. Once the plate 
information is available in digital form, the aging process of the photographic emulsion, which 
inevitably takes place during the long term storage in plate archives, is practically terminated.
If not cataloged and digitized, large quantities of astronomic photographic plates and films 
stored worldwide will be ultimately lost. 

Given the availability of suitable scanning and storage technology, projects to digitize 
the plates and to make the digital data available in the context of virtual observatories 
have been started in many countries. Leading in this context is the Harvard DASCH project 
\citep{dasch1}, many other observatories are now following the approach to digitize their 
plate collections, with many of them using commercial flatbed scanners of 
type Epson EXPRESSION 10000 XL.

Hamburger Sternwarte, an observatory inaugurated in 1912 in Hamburg-Ber\-ge\-dorf, was 
designed for state-of-the-art research. At the time of its construction, all new telescopes were 
already equipped with cameras and photographic plates as detectors. For some telescopes 
objective prisms were available to accommodate the upcoming need for spectroscopic 
astrophysical research. During more than 80 years of observations in Hamburg, astronomical plates were exposed and most of them are still stored and available in 
the observatory's plate archives. These plates together with observers' notes and 
logbooks present a unique and when digitized also useful data base for modern long-term studies and demonstrate the development of astronomical 
and astrophysical research and observational techniques throughout the 20$^{th}$ century. 
In 2010 a pilot project was initiated to study the feasibility of a complete digitization 
and open access presentation of the data in the Internet. Already on November 1$^{st}$ 2011, 
the first 3\,700 scans of the very oldest plates were presented 
\footnote{\url{http://plate-archive.hs.uni-hamburg.de}\,(D.\,Groote)} 
and since then all scanned plates have been made publicly available immediately.

In 2012, a digitization project was started by a consortium of Potsdam, Bamberg and Hamburg observatories, funded by the Deutsche Forschungsgemeinschaft (DFG). The goal of the project is to digitize all of the nearly 100~000 plates of the three observatories, to publish the data in the APPLAUSE database \citep[\textbf{A}rchive of \textbf{P}hoto\-graphic \textbf{PL}ates for \textbf{A}stro\-nomical \textbf{USE}, 
see e.g.][]{prague1, prague2}, and to make the data also available through the German Astrophysical Virtual Observatory (GAVO).

At the time of writing,
more than 31\,000 plates have already been digitized in Hamburg and are available
for download in high resolution. The plates can be accessed through a web interface. 
Each photographic plate is presented on an individual web page along with the associated
metadata and scans of observer notes and logbook entries, a low resolution image,
and links to the high resolution scans. An easy-to-use archive search includes the emulsion types, 
telescope modes and filters as well as coordinate search and exposure times.

In parallel to the digitization process, a data analysis framework has been developed that allows 
to retrieve astrometric and photometric data from the plates. To demonstrate the scientific value of the plate archives, we developed calibration procedures to extract otherwise unavailable historical information on bright BL~Lac type active galactic nuclei, which 
were ``unknown'' objects when the plates were taken. Yet these sources are notoriously 
variable and long-term variability studies are useful to characterize the stochastic nature of 
their variability. Here we will briefly describe the overall digitization project in 
Sect.~\ref{project}, and continue with the description of the automated astrometric and the photometric
calibration in Sect.~\ref{analysis}. Our target selection and the results of the analysis of bright 
{BL Lac} type objects are given in Sect.~\ref{results} and discussed in Sect.~\ref{discussion}.\\

\section{Photographic plates}
\label{project}

Photographic plates have been used both as detector ${\it and}$ storage medium for astronomical
observations for more than a century. Although the original plates are degrading with
time due to environmental conditions, most of the information is still available, even for plates 
that are 100 years old. One can clearly see some darkening caused by the glue used for the 
envelopes containing the plates if the emulsion side of the plate was exposed to it for a long time. 
Additionally, a silvering of the surface is observed for the oldest plates. The actual loss of 
information cannot be quantified as there are no comparable plates without any aging effects.

Digital images obviously do not degrade. Still, accidental data loss can occur for digitized images by purging or by 
aging of storage media. At present all images are stored at three different locations. The original ones are stored
at Hamburger Sternwarte using a RAID6-server. A backup is stored on a tape-robot at the 20~km distant university, and additionally, the data are kept at Potsdam Observatory.
Nevertheless, in contrast to plates kept in a room, it needs additional care to safely preserve the digital data over decades.\\

\subsection{The content of the plate archives}

Hamburger Sternwarte has a stock of about 45\,000 photographic plates and films, exposed between 
1905 and 1998. They are, first, sorted by telescope and, second, by consecutive plate number 
which is equivalent to sorting by exposure date. In some cases they are additionally ordered 
by the size. The plates are kept in a climate-controlled room at constant 
temperature~(20\textdegree) and humidity~(50\%) to prevent excessive aging and an infection with mold. 

Hand-written material is available with additional information concerning the objects 
and the exposure. Further information on the 
instrumentation and observation is provided by the notes of the observer which were taken at 
the time of observation. For scientific use, the 
hand-written material is also digitized and related to the plate images.

\subsection{Scanning process}

Flatbed scanners of the type Epson EXPRESSION 10000XL are used for all scans. Hand-written 
material is scanned at 150\,dpi and in color. Plates and films are normally scanned as 
16 bit grayscale PNG-images at 2\,400~dpi. With this setting, about 100 MB for plate sizes of 6.5~cm~x~9~cm up to 1.5~GB for plate sizes of 30~cm~x~30~cm 
will be produced per plate. Obtaining two high resolution scans per plate, approximately 1~TB of data per 1\,000 plates are produced.
In the case of 
plates with annotations on their emulsion-free back side, we obtained low resolution color scans at 300 dpi and used the resulting images for the web pages.
Low resolution images are obtained by down-grading the high resolution scans, where such a scan was unnecessary. All logbooks and 
observer notes are scanned in advance to ensure the availability of the entire data set on the release of the plate scans. To counteract aging effects, plate scanning is started with the oldest plates of a given telescope.

\subsection{Web pages and archive search}

The data including the scanned images are made available through a PHP-driven web interface 
which allows search queries to the SQL-database; the search results are also available as Comma-Separated-Value (CSV) lists.
The archive can be searched for plates covering certain
coordinates either by specifying a search box or a circular region by defining a radius. Furthermore,
special telescope modes, emulsions or 
filters can be chosen from a \textit{pull-down} menu. Most of the data can also be 
used for exclusion.
The search results also include the low resolution plate images, envelope, 
observer notes of the night, logbook entries and links to the high resolution images 
in FITS-format, if available. Later, after finishing the APPLAUSE-project, these headers 
should also contain the plate astrometry.

\section{Plate analysis}\label{analysis}

\subsection{Source extraction and astrometric calibration}
\label{astrometry}

The extraction of sources from the plate scans and the calibration of the
source coordinates are carried out with the PyPlate
software\footnote{\texttt{http://www.plate-archive.org/pyplate/}}, developed in
the context of the APPLAUSE-project. Specifically, the following processing
steps are implemented in the PyPlate workflow.\\

\begin{figure}
\includegraphics[width=\linewidth]{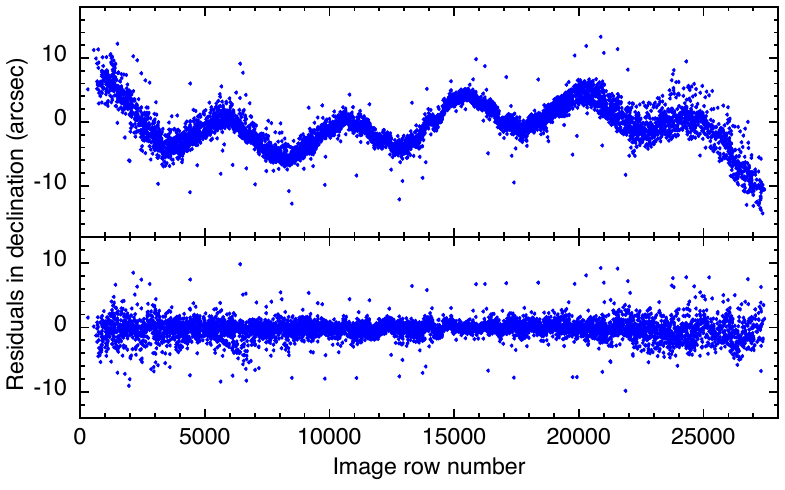}
\caption{Astrometric residuals in the plate LA00784 as a function of image row
number (Y-coordinate) after initial solution (top) and after refined solution
in sub-fields (bottom). The wobbles, introduced by a flat-bed scanner,
disappear after refinement, leaving residuals smaller than 5 arcsec for $\approx 99~\%$ of the objects
in the central part of the image. }
\label{figure_astrometry}
\end{figure}

The plate images are first analyzed with the SExtractor program \citep{Bertin1996} that indentifies all 
sources above a specified brightness threshold. We provide SExtractor with an inverted image 
and run the program in CCD mode. For each of the detected sources several parameters are determined 
and stored, of which the most relevant for the current work are the source centroid coordinates 
(\texttt{X}, \texttt{Y} in the image) and an elliptical-aperture magnitude (\texttt{MAG\_\-AUTO}).\\
An initial astrometric solution for the plate is computed with the Astrometry.net software \citep{Lang2010}. 
A list of coordinates of the brightest sources is used as an input and the Tycho-2 catalog is used as 
a reference. The Astrometry.net software then solves for the plate coordinates and provides a 
World Coordinate System (WCS) data structure as an output, which can be used for transforming 
image coordinates into celestial coordinates.\\
For refined coordinates, the image and associated object list is
recursively divided into two-by-two subfields. In each sub-field, the
astrometric solution is computed with the program SCAMP \citep{Bertin2006},
using the initial astrometric solution as a starting point and the UCAC4
\citep{UCAC401} astrometric catalog as a reference. The improved solution in a
sub-field is then used 
to transform the source coordinates into right ascension and declination. The
recursion is stopped when either the number of sources in the sub-field drops
below a specified threshold, 
or when the astrometric residuals estimated by SCAMP exceed the residuals in the
parent field, or when maximum recursion depth is reached. The described recursive solution helps 
eliminating complex image distortions like wobbles introduced by the flat-bed scanner; 
as an example of such a case we show the distribution of the 
astrometric residuals before and after our correction in Fig.~\ref{figure_astrometry}. 
As demonstrated here, the correction procedure 
leaves residuals of less than 5 arcsec for $\approx$99~\% of the objects in the central region of the plate.
The two orthogonal scans of each plate could in principle be used as well to eliminate the
scanner related non-linearities. The procedure outlined above reaches already a sufficient accuracy while avoiding
the additional complication of cross-matching two different plate scans.

\subsection{Photometric calibration}
\label{photometry}

As a result of the astrometric calibration the coordinates as well as the appropriate scanner magnitude of all objects found on the plate are available. We now continue to determine the correlation between the 
apparent stellar brightness and the measured scanner magnitude of an object, estimated by 
SExtractor (see~\ref{astrometry}). To this end, we match the coordinates given by the astrometric solution with the optical source catalog UCAC4. A match is accepted if 
the angular separation between the plate source and the catalog position is less than 5~arcsec. 

\begin{figure}
\includegraphics[width=\linewidth]{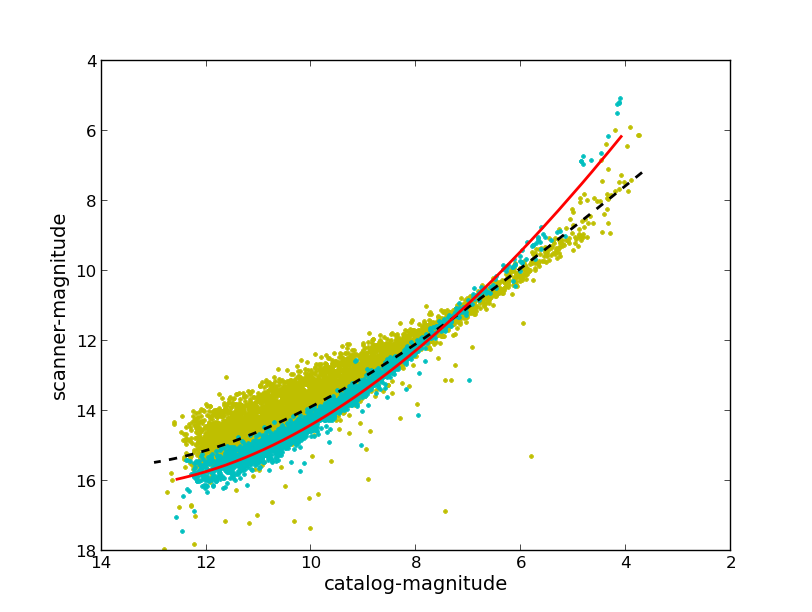}
\caption{Characteristic curve (black dashed line) of plate LA00784 calculated with all objects 
(yellow data points) and characteristic curve (red line) calculated with objects within a circle 
with a 6\,000 pixel (2.4\textdegree) radius around the center (pale blue data points).}
\label{figure_charcurve}
\end{figure}

In order to illustrate the problems encountered we show a comparison between 
instrumental and catalog magnitudes in Fig.~\ref{figure_charcurve}
for the plate LA00784 which covers a region of $11.25^\circ \times 11.25^\circ$. A variety of effects influences the accuracy of the derived 
instrumental magnitudes, some of them introduced by the telescope, of which comatic aberration seems to be the dominant one in the case of parabolic mirrors, but vignetting in the case of astrographs which were used in the present case. Therefore, with increasing distance to the center of the plate, the point spread function becomes less symmetric, which imposes systematic errors on the determination of instrumental magnitudes.
Figure~\ref{figure_charcurve} shows the relation between scanner magnitude and cataloged B magnitude from the UCAC4 for sources from the entire plate (yellow dots, dashed black line) and a 2.4\textdegree~diameter circular area around the center (blue dots, solid red line). The lines represent 4$^{th}$ order polynomial regression curves.
Clearly both the shape of the curves and the dispersion around the regression lines change.
In our analysis, we determine estimates of the characteristic curves from the central portion of the plate.
In particular, only sources within a circle around the plate center were considered; the radii for the individual plates are given in Table~\ref{tab:plates}.

\begin{figure}
\includegraphics[width=\linewidth]{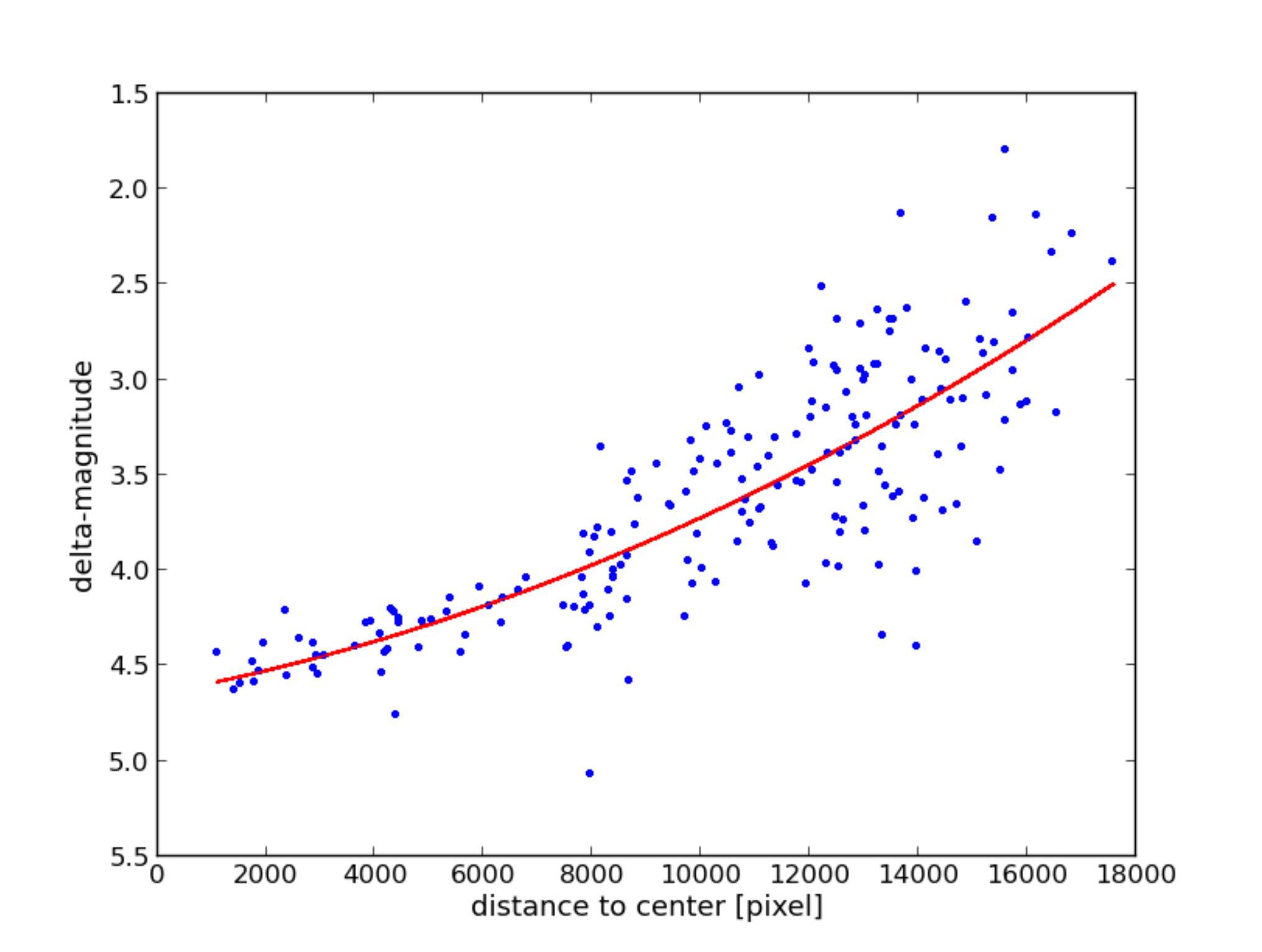}
\caption{Difference between scanner and catalog magnitude ($\Delta$-magnitude) for all objects on plate LA00784 with catalog magnitudes between 13.5~mag and 13.6~mag as a function of distance to the center.}
\label{figure_deltamag}
\end{figure}

The effect of vignetting on this characteristic curve is demonstrated for plate LA00748 in Fig.~\ref{figure_deltamag}, where the 
difference between measured scanner magnitude and the catalogue magnitude is shown as a function 
of distance to center (in pixels) for objects from the photographic plate with a catalog magnitude in 
the interval of 13.5~mag and 13.6~mag. As is clear from Fig.~\ref{figure_deltamag}, the further away from the 
plate center an object is located, the fainter its derived scanner magnitude becomes, although the intrinsic 
magnitude is the same; in the example shown, the difference is about two magnitudes.\\

Other reasons for the seemingly large scatter of the derived magnitudes can be attributed to both measurement errors
and plate errors such as impurities or scratches as well as to the inclusion of intrinsically variable objects in the analysis. Further, the scanner magnitude on the plate depends on 
the emulsion type and its wavelength dependent sensitivity through the color of objects.
Catalog data offer the possibility to choose between different color bands. As a result of the blue emulsion
types of the plates, we found that B-band colors result in the smallest scatter for the characteristic curve.
The remaining systematic error estimated by the standard deviation from the plotted curve is less than 
half a magnitude for most plates.

The statistical error is given by the standard deviation of measurements around the regression line in
a $\pm 0.1$~mag wide interval around the resulting catalog magnitude of the source. Having calculated the 1-$ \sigma $ confidence level of the scanner magnitude,
an apparent brightness along with an error can now be assigned to each scanner magnitude by the characteristic curve.\\

\subsection{Determination of limiting magnitudes}
\label{depth}

The photographic plates have a finite limiting sensitivity which depends to a large extent on 
the exposure time but also on the emulsion type used. To determine the lower brightness limit 
at which objects can still be detected on the plates, we investigate the completeness of objects 
found on the plate by comparing with more sensitive, recent catalogs; we note the additional complication that the limiting sensitivity depends on the
position on the plate.\\

\begin{figure}
\includegraphics[width=\linewidth,height=120mm]{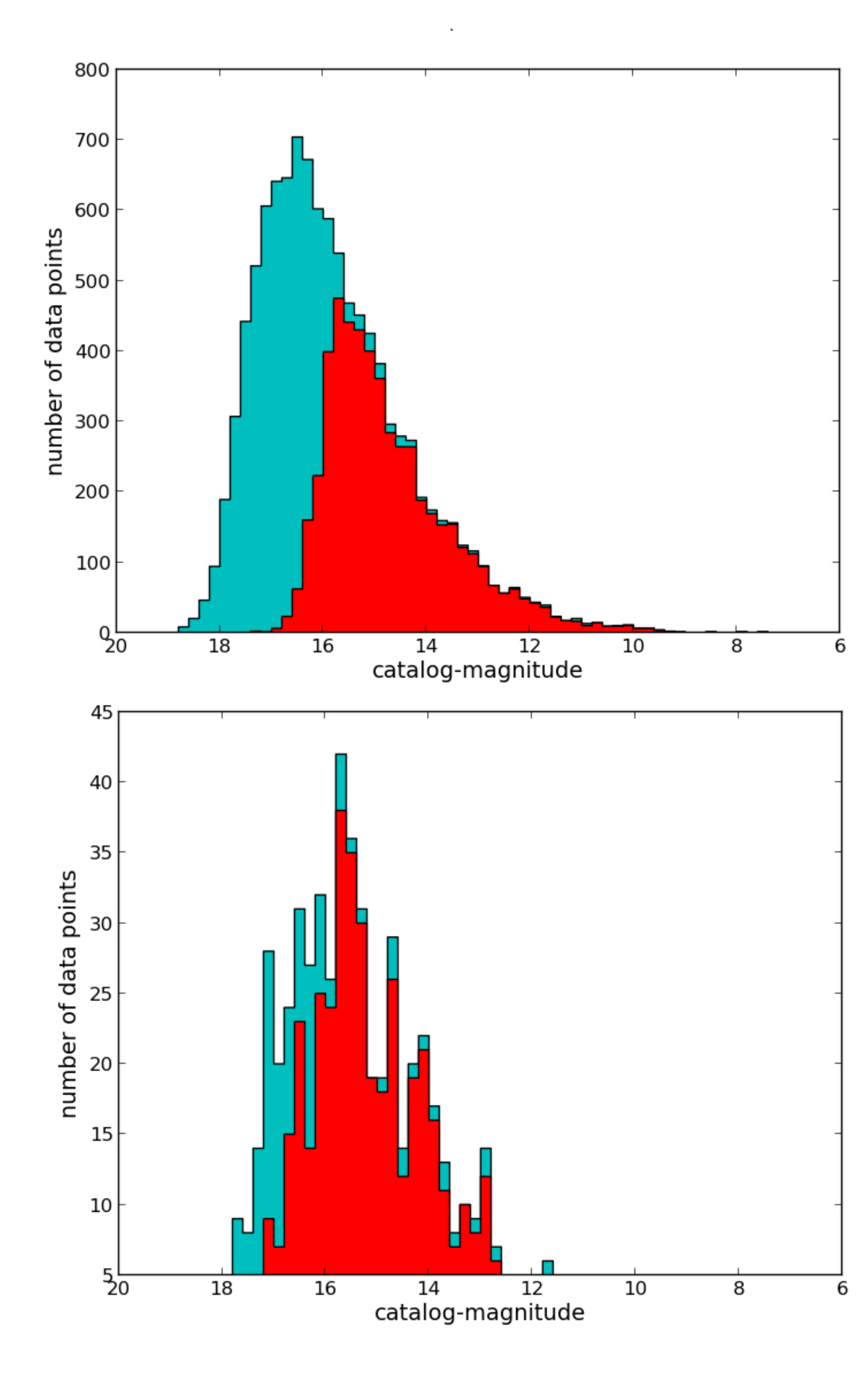}
\caption{Distribution of data points detected in a 1\textdegree~radius (red) and distribution of 
catalog objects in the same region (pale blue) dependent on their magnitude (top panel: 
plate LA06416, bottom panel: plate LA01609).}
\label{figure_limit}
\end{figure}

\begin{figure}
\includegraphics[width=\linewidth,height=60mm]{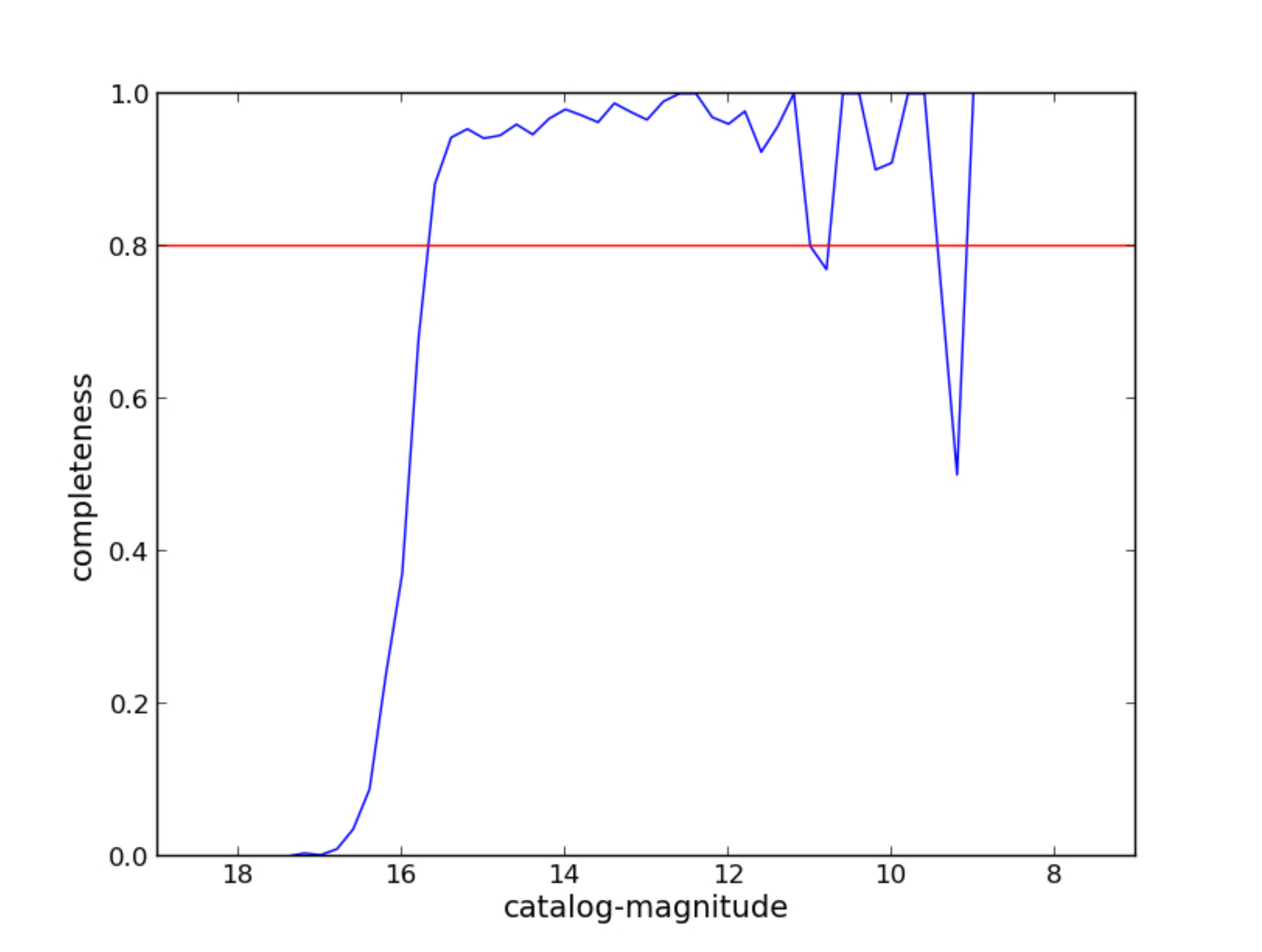}
\caption{Completeness of the photographic plate given as the relation of the distributions in Fig.~\ref{figure_limit} (top panel). 
The red line defines a completeness of 80\%. For this example the limiting magnitude is 15.8 mag.}
\label{figure_complete}
\end{figure}

To determine the limiting magnitude of a plate, we compare -- in a chosen section of the plate -- the distribution of
catalog magnitudes of detected sources with the magnitude distribution of all cataloged objects. An example of this procedure is shown in Fig.~\ref{figure_limit}, where the plate sources are represented
by the blue histogram and the cataloged sources
by the red histogram.
The top panel shows the distributions for plate LA06416, which clearly demonstrates that the catalog is
deeper, i.e., more complete, for fainter stars.  While in such cases a clear magnitude limit can be
defined, representing the limiting magnitude of the plate, this is hardly possible for plates showing
sections of the sky with low stellar density (see bottom panel).
In Fig.~\ref{figure_complete} we show 
the fraction of objects detected on the plate as a function of magnitude. For the brightest objects, the 
detection fraction is high. 
Fluctuations in this region can be attributed to inaccuracies in the astrometric solution, and 
individual mismatches produce large variations as a result of small number statistics. When inspecting the data carefully, a bright star of ninth magnitude was missed in the matched 
object list.  This missing cross-match of UCAC4 676-075960 is the result of  two 
closeby bright objects (HD 189376 and HD 189375) which are separated by 17 arcsec and can not be matched with the
single entry in the UCAC4 catalogue.
For fainter sources, a 
detection threshold can be defined. We consider a detection fraction of at least 80~\% necessary
to achieve reliable source coverage and define the 
limiting magnitude accordingly. Using this method, we can also provide a limit on the apparent 
brightness for a non-detected object.\\

\section{Target selection and scientific background}\label{results}

The historical data made available provide many unique opportunities to investigate 
the past properties of celestial objects, in particular of objects that were ``unknown'' at the time.
To demonstrate the scientific usefulness of the created data base we recover historical 
light curves of active galactic nuclei (AGN) observed serendipitously on some plates. 
Objects of the BL~Lac type class of AGN are notoriously variable and of particular interest as the emission is 
believed to be dominated by the relativistic and collimated outflow of their jets which for these objects are aligned
close to the line of sight. Historical data can considerably increase the range of time scales probed by measurements.

Studies carried out on time scales of up to 10 years indicate that these objects tend to have a 
different variability pattern for longer time scales, which is characterized by their power spectral density turning over to white noise \citep{kastend}.
Due to their large temporal coverage, the photographic plates have the potential to add substantial information to the time series analysis. Along a sample of $\gamma$-ray loud BL~Lac type objects, 
which are known to be bright in the optical region, we have identified a number of plates, which cover the position of the objects 1ES~1215+303 \citep{Aleksic01} and 3C~273 \citep{3C273}. These plates have been analyzed using the methods described in section~\ref{photometry} and \ref{depth} to extract light curves.

\subsection{1ES 1215+303}
\begin{figure}
\includegraphics[width=\linewidth]{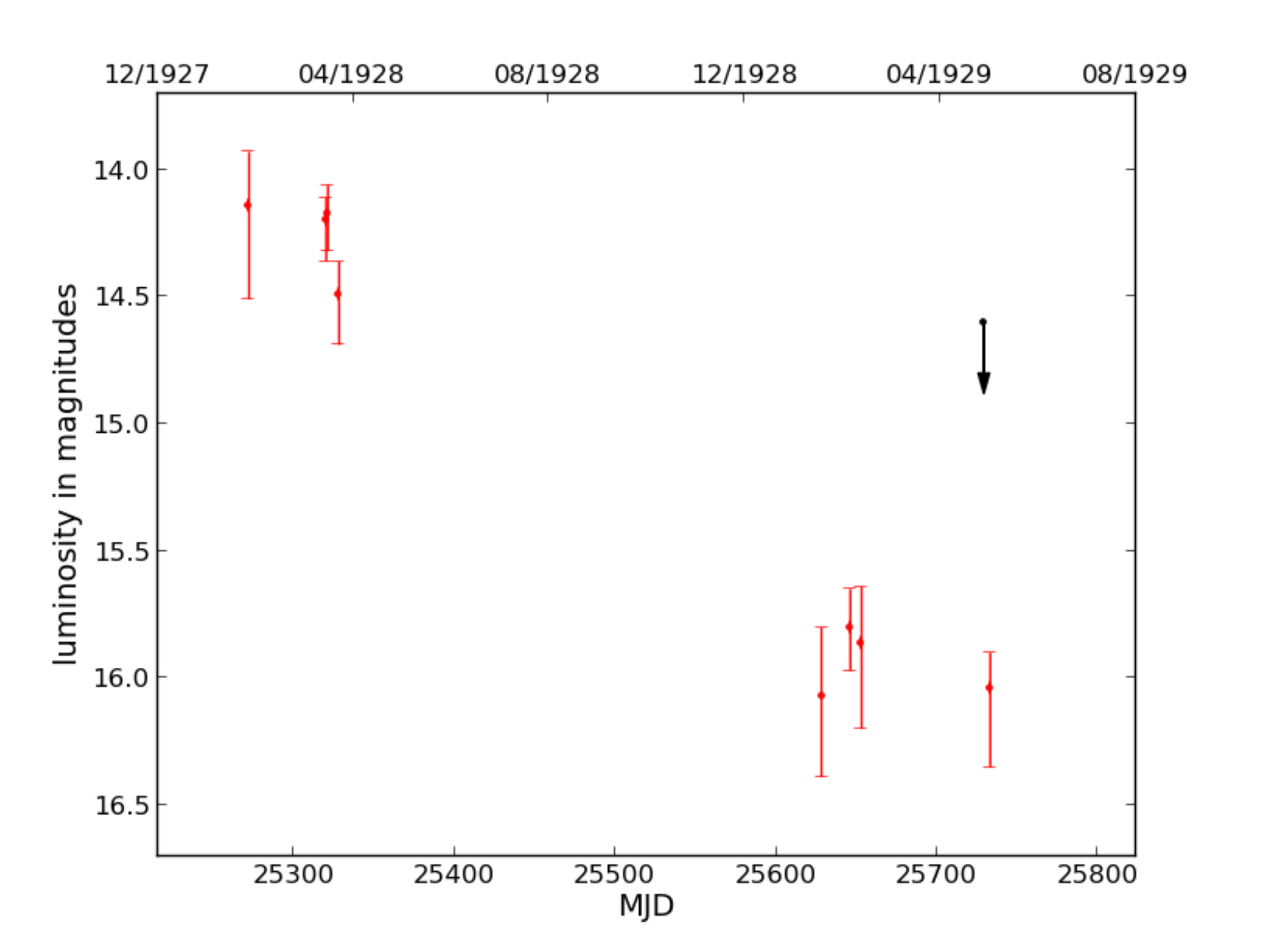}
\caption{Light curve of blazar 1ES~1215+303 including eight data points and one upper limit.}
\label{figure_lc_1ES}
\end{figure}

The brightness of this object has been 
determined for the period from January~1928 to May~1929 including eight data points and one upper limit; the derived 
light curve is shown in Fig.~\ref{figure_lc_1ES} with a time scale in MJD. There is a large drop in brightness 
from $14.25^{+0.07}_{-0.12}$~mag, the mean value of the first four data points, to $15.94^{+0.09}_{-0.13}$~mag 
(mean value of the last four magnitudes) within only about 300 days. The upper limit is consistent with the adjacent data points. Compared to the flux drop, the individual errors with a size between 0.1 mag 
and nearly 0.4 mag are small.
The brightness
at the first 1928 epoch is consistent with the more recent R-band measurements presented by
\citet{Aliu2013}, during which, however, no such excursion was observed.
The light curve extracted from these
over 80 year old plates clearly demonstrates that
1ES~1215+303 has shown at least one episode of strong optical variability. The timescale has still
to be determined.

\subsection{3C 273}

\begin{figure*}
\includegraphics[width=\linewidth]{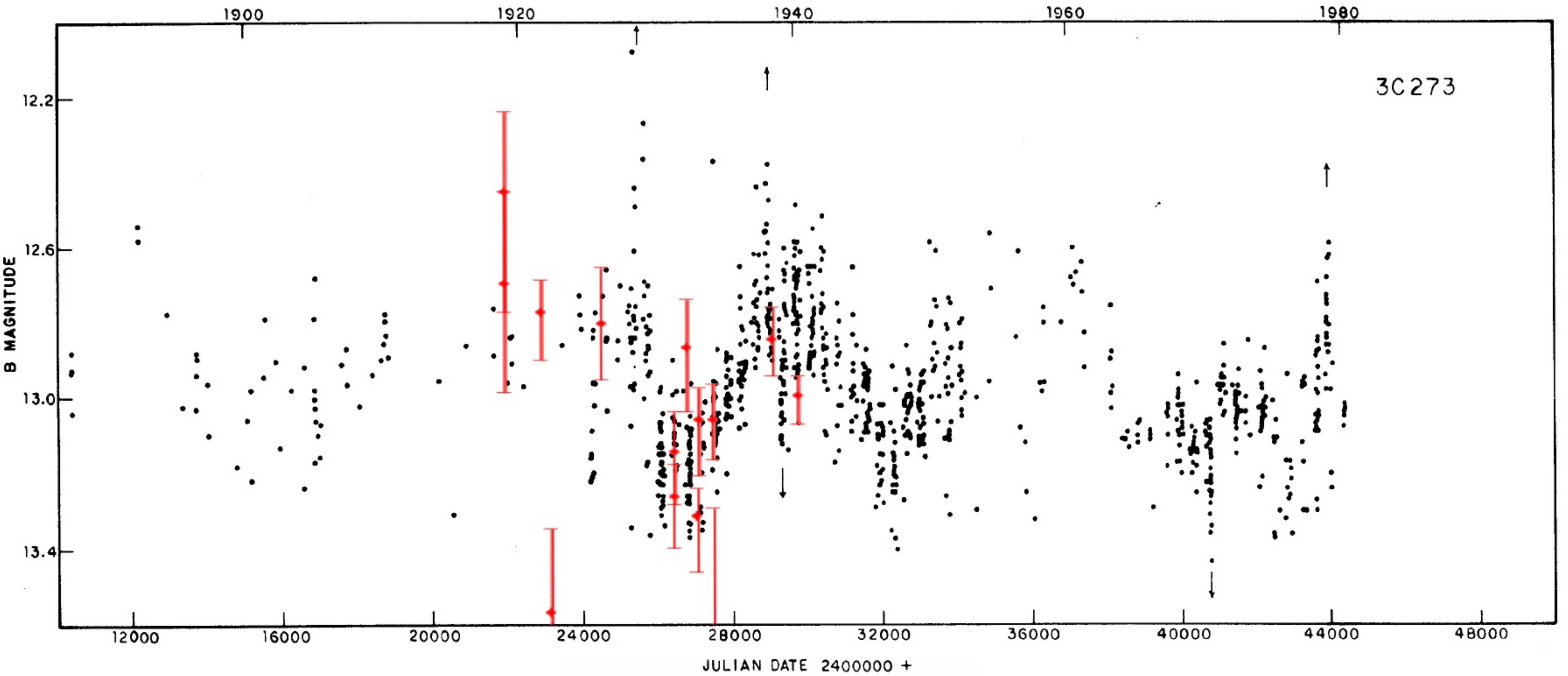}
\caption{Historical light curve of the quasar 3C~273 from \citet{Angione01}(black data points) together with our new data points (shown in red).
}
\label{figure_lc_3C}
\end{figure*}

Another light curve compiled and compared with historical data is that of the quasar 3C~273, which was 
detected on 15 plates re\-corded in the period from April~1916 to April~1940. Its light curve is displayed in Fig.~\ref{figure_lc_3C}, superimposed on a published light curve \citep{Angione01} covering a period of over 90~years. The additional 
data overlap and coincide with available archival data and thus provide an additional cross check that the calibration 
methods for stellar objects works equally well for AGN which usually have a different type of continuum spectrum than 
stars used for the photometric calibration.\\

\section{Summary and discussion}
\label{discussion}

The progress of the APPLAUSE project demonstrates that the content of the 
plate archives at Hamburger Sternwarte can be digitized, accessed through the Internet and utilized for 
scientific investigations. The digital data archive 
recovers a huge amount of material which can be used for scientific analysis.
We estimate that a historical record of positions and magnitudes of about two billions 
of objects is now easily accessible to the scientific community. The photometric calibration of the data 
compares well with other, similar efforts and yields accuracies of typically 0.2-0.5 mag. 
In addition, a large number of spectra, normally taken with an objective prism, are also available on the plates 
of the Hamburger Sternwarte plate archives. While these spectra normally have relatively low resolution and can 
therefore be easily superseded by modern spectra, a number of historic spectra exist, for example of the NOVA GEM 1912.
These spectra can now be extracted and used for modeling with state-of-the-art nova modeling tools.
However, quite naturally, most of the work possible with the plate archives deals with light curves.
\\
Our analysis of the plates stored in the plate archives of Hamburger Sternwarte shows that photographic plates 
can still be used effectively with modern data analysis, which we have shown to provide results consistent with previous findings. The research of highly variable objects like blazars 
or qua\-sars using these historical data allow glimpses into the past that would be otherwise impossible. 
Conclusions about the physical conditions can be drawn from fluctuations of the resulting long-time light curves, 
and photographic plates are the only medium to access recordings that date back so far.
\\
Clearly, the analysis carried out, so far, covers only a fraction of the available potential;
for example, only for about 50\% of the plates, the coordinates of the plate centers are listed in the metadata, 
while the rest is usually annotated by the name of the sky region observed, for example, ``Pleiades'',
``Kom. 1912d Westphal'' 
or similar, or is just empty. Normally the plate center of such plates can be reconstructed with
high accuracy using the ``astronomy.net'' software. We expect that the full archive will be
completed and available by mid 2017.

\newpage

\acknowledgements
The digitization of Hamburg plate archives is supported by the Deutsche
Forschungsgemeinschaft (DFG) grant GR 969/4-1. T.
Tuvikene acknowledges financial support by the DFG grant EN 926/3-1 and by the Estonian Research Council grant PUTJD5.
This work was only possible due to the painstaking work of our scanning personnel (Polzin, K., M\"uller, A.,
Preller, S., Vollersen, A.) and honorary helpers (Engelhardt, N., Gabriel, H., Wulff, A.)

\bibliographystyle{an}
\bibliography{bib}

\newpage
\appendix

\section{Appendix}

\renewcommand{\arraystretch}{1,4}
\begin{table*}
\small
\caption{Listing of analyzed plates including extracted B magnitude or limiting magnitude (lim), date of recording, radius of chosen circle, exposure time and emulsion type.}
\label{tab:plates}
\begin{tabular}{|l|c|c|c|c|c|c|}
\hline
\textbf{object} & \textbf{B magnitude} & \textbf{date [YYYY-MM-DD]} & \textbf{platenumber} & \textbf{radius [pixel]} & \textbf{exposure time [min]} & \textbf{emulsion} \\ \hline
3C 273 & $12.42^{+0.20}_{-0.30}$ & 1916-04-29 & LA00418 & 9000 & 61 & Agfa isorapid\\ 
 &  $12,65^{+0.23}_{-0.27}$ & 1916-05-02 & LA00428 & 10000 & 61 & Agfa isorapid\\ 
 &  $12.72^{+0.08}_{-0.12}$ & 1919-04-25 & LA00651 & 8000 & 61 & Agfa isorapid\\ 
 &  $13.47^{+0.21}_{-0.30}$ & 1920-03-09 & LA00773 & 8000 & 60 & Schleussner\\ 
 &  $12.75^{+0.14}_{-0.14}$ & 1924-03-13 & LA01366 & 10000 & 240 & - \\ 
 &  $13.07^{+0.10}_{-0.13}$ & 1930-02-24 & LA02964 & 9000 & 240 & Agfa Astro Z\\ 
 &  $13.18^{+0.08}_{-0.13}$ & 1930-02-25 & LA02971 & 9000 & 120 & Agfa Astro Z\\
 &  $12.81^{+0.12}_{-0.16}$ & 1931-03-15 & LA03306 & 10000 & 240 & Agfa Astro Z\\ 
 &  $13.23^{+0.07}_{-0.14}$ & 1932-02-05 & LA03595 & 8000 & 180 & Agfa Astro V\\ 
 &  $12.99^{+0.08}_{-0.14}$ & 1932-02-29 & LA03635 & 9000 & 240 & Agfa Astro V\\ 
 &  $12.99^{+0.09}_{-0.10}$ & 1933-04-26 & LA04094 & 9000 & 200 & Agfa Astro Z\\ 
 &  $13.59^{+0.38}_{-0.42}$ & 1933-05-28 & LA04118 & 9000 & 34 & - \\ 
 &  $12.79^{+0.08}_{-0.09}$ & 1938-03-20 & LA05295 & 8500 & 50 & Agfa Astro Z\\
 &  $12.93^{+0.05}_{-0.07}$ & 1940-04-08 & LA05852 & 9000 & 60 & Agfa Astro Z\\ \hline
   1ES 1215+303 & $14.14^{+0.21}_{-0.37}$ & 1928-01-27 & LA02183 & 6000 & 20 & - \\ 
    & $14.20^{+0.09}_{-0.16}$ & 1928-03-15 & LA02266 & 7000 & 192 & Agfa isorapid\\ 
    & $14.17^{+0.11}_{-0.15}$ & 1928-03-16 & LA02275 & 10000 & 180 & Agfa isorapid\\ 
    & $14.49^{+0.13}_{-0.20}$ & 1928-03-23 & LA02291 & 5000 & 46 & Agfa isorapid\\ 
    & $16.07^{+0.27}_{-0.32}$ & 1929-01-17 & LA02503 & 4000 & 36 & Agfa isorapid\\ 
    & $15.80^{+0.15}_{-0.17}$ & 1929-02-04 & LA02527 & 4000 & 240 & Agfa Astro V\\ 
    & $15.86^{+0.22}_{-0.35}$ & 1929-02-11 & LA02544 & 5000 & 120 & Agfa Astro V\\ 
    & $16.04^{+0.14}_{-0.31}$ & 1929-05-02 & LA02595 & 3500 & 180 & Agfa Astro Z\\ 
 & 14.6 (lim) & 1929-04-28 & LA02592 & - & 20 & - \\ \hline
\end{tabular}
\label{Ergebnisse1}
\end{table*}

\end{document}